# Direct in-situ measurement of electrical properties of solid electrolyte interphase on lithium metal anode

Yaobin Xu[1†], Hao Jia[2†], Peiyuan Gao[3], Diego E. Galvez-Aranda[4,5], Saul Perez Beltran[4], Xia Cao[2], Phung M. L. Le[2], Jianfang Liu[6], Mark H Engelhard[1], Shuang Li[1], Gang Ren[6], Jorge M. Seminario[4,5,7]*, Perla B. Balbuena[4,7,8]*, Ji-Guang Zhang[2], Wu Xu[2]*, Chongmin Wang[1]*

[1]Environmental Molecular Sciences Laboratory, Pacific Northwest National Laboratory, Richland, WA 99354, USA

[2]Energy and Environment Directorate, Pacific Northwest National Laboratory, Richland, WA 99354, USA

[3]Physical and Computational Sciences Directorate, Pacific Northwest National Laboratory, Richland, WA 99354, USA

[4]Department of Chemical Engineering, Texas A&M University, College Station, TX 77843, USA

[5]Department of Electrical and Computer Engineering, Texas A&M University, College Station, TX 77843, USA

[6]The Molecular Foundry, Lawrence Berkeley National Laboratory, Berkeley, CA 94720, USA

[7]Department of Materials Science and Engineering, Texas A&M University, College Station, TX 77843, USA

[8]Department of Chemistry, Texas A&M University, College Station, TX 77843, USA

†These authors contributed equally to this work

*Corresponding authors. Email: seminario@mail.che.tamu.edu, balbuena@mail.che.tamu.edu, wu.xu@pnnl.gov, chongmin.wang@pnnl.gov

**Solid electrolyte interphase (SEI), a thin layer that dynamically forms between active electrode and electrolyte during battery operation, critically governs the performance of rechargeable batteries[1-5]. An ideal SEI is expected to be electrically insulative to prevent persistently parasitic reactions between the electrode and the electrolyte, while ionically conductive to facilitate Faradaic reactions of the electrode[1,2,6]. However, the true nature of the electrical properties of an SEI layer remains hitherto unclear due to the lack of a direct characterization method, leaving a range of behaviors of rechargeable batteries unelucidated. Here, we use in-situ bias transmission electron microscopy, for the first time, to directly measure the electrical properties of SEIs formed on copper (Cu) and lithium (Li) substrates. Surprisingly, we discover that, in terms of electrical behavior, SEI is distinctively different from a typical electrical insulator as what has been widely, and up to date, assumed ever since the discovery of SEI; rather, SEI shows voltage-dependent differential**

**conductance. We reveal that the thickness and the topographic feature of SEI, which have been regarded as the critical factors affecting the electrochemical properties, are essentially controlled by the rate of differential conductance of SEI layer. A higher rate of differential conductance induces a thicker SEI with intricate topographic feature, consequently leading to an inferior Coulombic efficiency and cycling stability. We illustrate that SEI components with high content of inorganic species show high electron resistance, while those in rich of organic species exhibit higher electron leakage, which are corroborated by the results of reactive force field and hybrid ab initio molecular dynamics simulations of SEI and correspondingly electron transport calculations. The work solves the long-standing mystery as how SEI functions electrically during the electrochemical operation, unveiling that the rate of differential conductance of SEI layer is deemed to be the root cause that governs characteristics of both SEI and Li deposit and correspondingly the battery performance. The work provides unprecedented insight for targeted design of SEI with desired characteristics towards better battery performance.**

Functioning of an electrochemical cell, typically such as a rechargeable battery, depends on the synergy of three major components in the cell: anode, electrolyte and cathode. The electrolyte, in either solid state or liquid state, is sandwiched between the cathode and the anode to facilitate ion transport[5,7-11]. The interface between electrolyte and electrode is not atomically sharp, instead electron transfer across the interface leads to the formation of an interphasial layer, which is termed as solid electrolyte interphase (SEI) layer[1-5]. It is the characteristics of the SEI layer, including chemical, structural, morphological and mechanical properties[5,12-16], that determine a series of key properties of rechargeable batteries, such as loss of active ion inventory, cycle life, rate capability, and high/low temperature performance of a rechargeable battery[2,6,17-21].

For better battery performances, the SEI is expected to possess three ideal characteristics: electrically insulative, ionically conductive, and constant thickness[2,3]. These three characteristics are interactively correlated, typically, the thickness of SEI layer is controlled by the electrical properties of SEI layer. However, SEI layers do not seem to behave the ideal characteristics, rather the thickness of the SEI continuously increases during charge-discharge cycling and shelf storage, indicating that the SEI does not behave as an electrical insulator[1,2,6,22-25]. Electronic structure calculations indicate that certain SEI components and their grain boundaries, in contrast with their crystalline counterparts, are prone to electron leakage through the SEI layer[26-32], leading to continued thickening of SEI layer. A thick SEI increases the ion conduction pathway and thus the ionic resistance, deteriorating the kinetic properties of batteries[6,19,27,33]. In addition, the SEI growth also consumes active ion source and electrolyte in the batteries, leading to capacity decay, short cycle life and calendar life of the batteries.

Despite the critical importance of the electrical properties of SEIs, quantitative measurement of this parameter remains unsolved due to the lack of a proper and reliable method. The four-point Hebb-Wagner polarization and electrochemical impedance spectroscopy (EIS) methods for mixed ionic-electronic conductors cannot be readily applied to quantify the electrical conductivity of SEIs[34-39], because SEIs are not only highly air-sensitive but also very thin that is beyond the high spatial resolution of the method. Scanning probe microscopy inside a glove box or scanning electron microscopy could solve the air-sensitive issue[40-48]. However, the scanning probe method is based on the principle of touching the top surface of the sample, while without any information from the top surface to the counter electrode. Adhering to the nature of this limitation, the thickness of the SEI layer at the measuring site cannot be in-situ measured. Consequently, it is hard to directly correlate microstructure and chemical information of SEI with the measured resistivity. In spite of the lack of concrete experimental evidence, it is widely assumed that an SEI layer behaves as an insulator, as such assumption helps to interpret, at some degrees, the electrochemical performances of rechargeable batteries[49]. In essence, for all types of cell chemistries that are enabled by SEI, the electrical and ionic properties of SEIs remain as the most challenging mystery, leading to a range of behaviors of rechargeable battery being uninterpreted.

Here, we describe a novel in-situ bias transmission electron microscopy (TEM) approach to directly measure the electrical properties of SEI layers grown on copper (Cu) and lithium (Li) substrates, for the first time, revealing the electrical characteristics of SEI in terms of current (I)-voltage (V) relationship, differential electrical conductance, critical field strength and band gap. We unveil that the I-V characteristics of SEIs resemble certain electrical conductance, rather than electrical insulator as historically assumed. The SEI with a higher rate differential conductance tends to exhibit a greater thickness and more complex topographic features, consequently leading to an inferior electrochemical performance. The work highlights the governing role of electrical properties of the SEI layer and their tuning towards the enhanced performance of an electrochemical cell.

We integrated in-situ TEM with scanning tunneling microscopy (STM) technique to measure the electrical properties of SEIs on Cu and Li, as illustrated in Fig. 1 and detailed in the Supplementary Methods section and Supplementary Figs. 1-7. An STM tungsten (W) nanoprobe with atomically clean surface was used as the counter electrode (Supplementary Fig. 3), which was manipulated by the piezo-system with 3-axis nanometer scale control. As ion-blocking Cu and W electrodes are used, the measured I-V data directly reflect the electron transport behavior of SEI. It should be noted that this measurement with two blocking electrodes does not exactly resemble a real Wagner-Hebb polarization measurement, but rather provides an upper limit value of the electrical conductivity[8,34,50,51]. As SEI is very sensitive to electron beam[52-54], we performed the I-V measurements at very low magnification of electron dose rate of 1 e$^-$ Å$^{-2}$ s$^{-1}$ to avoid electron beam induced damage to SEIs, and did I-V curve measurement calibration to make sure the experimental results are repeatable and credible (Supplementary Figs. 8-14) To systematically study different SEIs, Li bis(fluorosulfonyl)imide (LiFSI) and 1,2-dimethoxyethane (DME) were chosen to make four electrolytes with designed microscopic solvation structures[55,56]: (1) a low concentration electrolyte (LCE) comprised of 1 M LiFSI in DME with a molar ratio of 1:9; (2) a high concentration electrolyte (HCE) of LiFSI and DME with a molar ratio of 1:1.2; (3) a localized high concentration electrolyte (LHCE) formed by adding bis(2,2,2-trifluoroethyl) ether (BTFE) diluent into the HCE to yield LiFSI-DME-BTFE=1:1.2:3 by mol. (LHCE-BTFE); and (4) an LHCE with bis(2,2,2-trifluoroethyl) ether carbonate (BTFEC): LiFSI-DME-BTFEC=1.0:1.2:3.0 by mol. (PLHCE, as free DME molecules are not closely coordinated with Li$^+$ and making it a pseudo-LHCE[56]) (Supplementary Tables 1-2).

The I-V curves of SEI layers formed on Cu and Li with the four different electrolytes are shown in Fig. 2. Due to the very thin nature of the SEI layer to nanoscale, prior to deciphering the physical meaning of the measured I-V curves of the SEI layers, we calibrated the measurement of the I-V curves with known materials at nanoscale as a standard. Therefore, we measured the I-V curves of SiO$_2$ as a typical insulator and TiO$_2$ as a semiconductor. As shown in Fig. 2a, even at nanoscale, the I-V curve of SiO$_2$ shows typical features of insulator, while that of TiO$_2$ is semiconductor. It is rather apparent that the I-V curves of SEIs on both Cu and Li are similar to that of TiO$_2$ but distinctive to that of SiO$_2$, revealing that the electrical properties of SEIs resemble that of a semiconductor. As detailed in the Method and Supplementary Information, based on the results of hybrid ab initio molecular dynamics (AIMD) simulation of SEI formation and the calculated electronic structure of these four electrolytes, we calculated the I-V curve for the SEI layer on Li-metal. The I-V curves (Fig. 2d) calculated using the Generalized Electron Nano-Interface Program (GENIP)[26,57-59] procedure exhibit similar characteristics and trends to those captured experimentally.

Two characteristic parameters can be extracted from the I-V curves to quantitatively interpret the I-V curves. One is the differential conductance, dI/dV, which is plotted as a function of applied voltage, V (Figs. 2e-h). Another one is the critical field strength for the breakdown of SEI layer. The differential conductance of all samples unanimously shows a linear relationship with the applied voltage. However, the slopes of the linear relationship, which can be termed as the rate of differential conductance, are significantly different for different samples. It would be expected that for an insulator, such as $SiO_2$, the dI/dV-V should have a slope of close to zero, which is consistently supported by what we have measured ($SiO_2$: $6.06 \times 10^{-27}$). For a semiconductor such as $TiO_2$, the dI/dV-V plot exhibits a positive slope ($2.19 \times 10^{-8}$). The differential conductance (dI/dV) of all SEIs on both Cu and Li shows linear positive correlations to the applied voltage, while the values of slopes follow a decreasing order from LCE ($3.86 \times 10^{-7}$ and $2.72 \times 10^{-7}$) to PLHCE ($1.22 \times 10^{-7}$ and $2.26 \times 10^{-7}$), HCE ($8.93 \times 10^{-8}$ and $2.53 \times 10^{-8}$) and LHCE ($7.67 \times 10^{-8}$ and $1.48 \times 10^{-8}$), where the values in the parentheses correspond to the slopes of dI/dV-V on Cu and Li, respectively. Since the differential conductance represents the electron density of state at the local position of the SEI layer, the positive linear relationship between dI/dV and voltage indicates that the electrical conductance increases with increasing voltage, implying that the formation of SEI during battery cycling shows dependence on the voltage difference between the electrode/SEI interface and the SEI/electrolyte interface. The larger the rate of the differential conductance against voltage is, the stronger the SEI responds to the voltage increase. As illustrated in Figs. 2f-g, regardless of the type of the substrate (Cu or Li), the SEIs formed by LHCE and HCE electrolytes show much lower rate of differential conductance than those by PLHCE and LCE electrolytes. The dI/dV-V plot derived from the calculated I-V curve (Fig. 2h) corroborate our experimental results. It should be noted that to account for the SEI layer thickness effect, we draw the differential conductance, dI/dV, as a function of the electrical field strength (voltage divided by the thickness of SEI layer) by which the SEI layer thickness effect is normalized as detailed in the Supplementary Information. As shown in Supplementary Fig. 15 (c), the electrical differential conductance against the electrical field strength shows similar trend of variation for the case of dI/dV as a function of V (Supplementary Fig. 15b).

With the increase of the voltage, the current increases parabolically and to a critical voltage, the current reaches a value that exceeds the maximum value of the instrument (Fig. 2). When applying constant voltage above the critical voltage, the current keeps saturated, indicating the transition from semiconductor to conductor is irreversible (Supplementary Fig. 7). Critical field strength is defined as the critical voltage divided by the sample thickness. The critical field strengths of the SEIs for the four electrolytes are different, which correlate positively with the slopes of the dI/dV-V plots as depicted in Figs. 2f-g. The critical field strength of SEI formed in LHCE is larger than those of SEIs formed in LCE and PLHCE, indicating the SEI formed in LHCE is much stable against increasing voltage as compared with those formed in other three electrolytes.

To demonstrate the direct correlation between SEI electrical property and battery performance, the electrochemical performances in terms of Coulombic efficiency (CE) and cycle life of those four electrolytes were evaluated in Li||Cu cells and Li||LiNi$_{0.8}$Mn$_{0.1}$Co$_{0.1}$O$_2$ (NMC811) batteries. As shown in Figs. 2i-j, the first cycle CEs of Li||Cu cells and the stable cycle numbers of Li||NMC811 cells have the following orders: LHCE > HCE > PLHCE > LCE (Supplementary Table 3). Overall, an increased differential conductance of SEI correlates to a decreased Li CE and battery cycling stability (Figs. 2i-j), indicating the governing role of SEI electrical property on the battery performance.

Consistent with above electrochemical property differences among these four electrolytes is the significant difference of morphological features of both SEI layer and the deposited Li. The deposited Li in these four electrolytes all exhibits crystalline structure and granular morphology (Supplementary Figs. 16-17). However, the particle size distributions and topographic features vary significantly. Figure 3a shows the morphologies of the deposited Li particles using high angle annular dark field imaging (HAADF) in scanning transmission electron microscope (STEM) by which the image intensity is proportional to the square of atomic number of the sample. The elemental compositions of SEI, such as O, C, F, S and N, each has large atomic number as compared with Li, leading to a large contrast between SEI layer and Li, therefore lending the convenience of delineating the spatial distribution of SEI layer. Based on the SEI layer configuration maps (Fig. 3b) derived from the STEM-HAADF images (Fig. 3a), it can be seen that SEI layer with a high rate of differential conductance is corresponded with a high SEI:Li metal ratio.

Three-dimensional (3D) visualization of Li deposits (Supplementary Video 1) yields details of Li topography. It is evident that for the SEI with a high rate of differential conductance and a low critical field strength, as representatively shown for the case of LCE (Fig. 3c), the deposited Li particles exhibit a wide size distribution, large fraction of isolated small particles (possible "dead" Li), and a high topographical tortuosity, leading to high specific surface area of the SEI. In contrast, for the SEI with a low rate of differential conductance and a high critical field strength, as represented by the case of LHCE (Fig. 3c), the deposited Li particles are large, uniformly distributed, and topographically smooth, leading to a low specific surface area of the SEI layer and less "dead" Li.

Thickening of SEI layer is a self-limiting process, which is governed by the electron leakage behavior the on-growing SEI layer. Our observations clearly indicate the SEIs formed on Cu (Supplementary Fig. 18) and Li (Fig. 4a) exhibit a similar trend of increasing thickness with a high rate of differential conductance and low critical field strength of the SEI (Fig. 4b). The SEI formed in LCE has the highest rate of differential conductance and the lowest critical field strength, which is corresponded with an SEI layer thickness of ~ 35 nm. The SEI layer formed in LHCE has the lowest rate of differential conductance and the highest critical field strength, corresponding to an SEI thickness of merely 7.5 nm.

Aiming to gain further insight on the origin of different electrical properties of different SEIs, the compositions of SEIs formed on Cu and Li were analyzed by cryo-TEM, energy-dispersive X-ray spectroscopy (EDS), electron energy loss spectroscopy (EELS), and X-ray photoelectron spectroscopy (XPS) as shown in Supplementary Figs. 19-28. Chemically, the SEI layer is composed of Li as the sole cation, which is balanced by the anions comprised of oxygen (O), sulfur (S), carbon (C), fluorine (F), and nitrogen (N). The SEI with a high O:S ratio tends to exhibit a high rate of differential conductance and a low critical field strength, whereas the SEI with a low O:S ratio leads to a low rate of differential conductance and a high critical field strength. The O:S ratios of the SEI layers on Cu and Li follow the order from high to low as LCE (4.91 and 19.81) > PLHCE (2.90 and 7.71) > HCE (0.92 and 1.58) > LHCE (0.78 and 0.69), where the values in the parentheses correspond to the O:S ratios of SEIs on Cu and Li, respectively. These values exactly follow the tendencies of gradually decreased rate of differential conductance and increased critical field strength (Supplementary Fig. 23). As discussed in detail in the Supplementary Information, the variation of O:S ratio represents the relative contribution of the solvent and salt anion derived components of SEI layer in these electrolytes. This observation clearly demonstrates that salt derived component in the SEI leads to low electrical conductance, while the SEI component derived from solvent yields high electrical conductance.

To delineate the critical factors, in particular molecular level information, that control electrical properties of SEI layer, we built Li-electrolyte interface models to investigate the SEI structure using hybrid AIMD-based simulation (Supplementary Figs. 29-40 and Tables 4-8) and subsequently calculate the electron transport in terms of I-V curve as representatively shown in Fig. 4d for the sampling SEI used for the I-V curve calculation. The concentrations of the various species in SEI derived based on hybrid AIMD generally agree with the XPS data (Supplementary Table 9). We found that SEIs formed in LCE and PLHCE, which exhibit high electrical conductance (Fig. 2), show greater proportion of organic to inorganic phase (higher C content as shown in Supplementary Fig. 40), indicating the dominance of solvent derived SEI components as what we have experimentally observed. The high proportion of organic components in SEI will lead to large porosity of SEI, presence of charged molecular fragments or organic radical species due to incomplete molecular reduction and existence of large amount of dissolved Li ions as a consequence of incomplete oxidation arising from the incomplete molecular reduction, which may lead to formation of "dead" Li as observed experimentally (Fig. 3c). All these collectively contribute to the electron leakage. It should be noticed that the less concentration of sulfate products in the calculated SEI, as contrasted with that captured from XPS, EDS and EELS, is attributed to the fast reaction rate between FSI$^-$ and Li metal and the difference between the electrolyte to anode (E/A) ratio, which may be lower in the simulations than in the experiment[60]. Indeed, we carried out additional simulation with high E/A ratio (Supplementary Figs. 35-37), which demonstrates increased sulfate products and the calculated I-V shows similar trend (Supplementary Fig. 38b).

Band gap is a parameter to reflect electron transition from valance to conduction band, which correlates with the electron tunneling barrier of the SEI layer[27,28]. To better understand the electrical properties of SEIs, we measured the band gaps of SEI layers using EELS (Supplementary Fig. 41)[7,61,62]. The band gap of the SEI on Li deposit shows two significant features (Fig. 4c). First, the average band gap of SEIs follows the increasing order as LCE (1.63 ± 0.12 eV) < PLHCE (1.86 ± 0.13 eV) < HCE (2.03 ± 0.19 eV) < LHCE (2.35 ± 0.14 eV), which corresponds well to the orders of increasing critical field strength and decreasing rate of differential conductance. Second, the band gaps of SEI layers in these four electrolytes show spatial variance from outer to inner SEI with exception of LHCE showing nearly a constant band gap value across the SEI. It is apparent that the spatial change of band gap across the SEI layer thickness direction correlates with the chemical composition variations of the SEI. As shown in Supplementary Fig. 24, the SEI layer formed in HCE shows a distinctive bilayer structure, where the outer layer has high F intensity, while O distributes near the Li deposit, and very few C distributes on the surface of the SEI layer. The composition difference across the SEI layer indicates the difference of electronic environment between the outer layer and the inner layer, hence the difference of band gap. The O-rich nature in SEIs consistently accounts for the formation of $Li_2O$ particles in the SEIs in LCE, PLHCE, and HCE, while SEI layers formed in HCE and LHCE contain S based components. Based on band gap calculation of SEI components as shown in Supplementary Fig. 42 and previous studies[29,30,63,64], it has been indicated that the electron leakage resistance of amorphous $Li_2S$ (3.07 eV) is higher than that of $Li_2O$ (2.2 eV)[64]. Furthermore, grain boundaries of inorganic compounds have been predicted to enhance electron tunneling in SEI layer[29,30].

It should be noted that the classic model of SEI is composed of an inner layer of inorganic and an out layer of organic[2], which, in terms of electrical properties, correspond to a tandem structure. Apparently, it would be expected that the inner layer of inorganic will be the determining layer on the electrical properties. However, given the fact that the inner layer of SEI is a composite structure with crystalline particles dispersed in the amorphous matrix (Fig. 4a and Supplementary Fig. 19), the electron leakage characteristic will be determined by the continuous amorphous matrix, rather than the dispersed crystalline particles.

An ideal SEI is considered to be highly ionically conductive while electrically insulative. Our direct measurement of the electrical properties of SEIs reveals the electrical behaviors of SEIs formed in four typical electrolytes. Contrary to what has been conventionally assumed, SEIs do not act as perfect electrical insulators, rather show non-negligible electrical conductance, which, as a root cause, governs the SEI layer formation and Li deposition and consequently affecting battery performance. A higher electrical conductance of SEI could facilitate electron transport inside SEI, especially at the initial stage of SEI formation, leading to reduction of $Li^+$ in SEI and consequently formation of metallic Li inside the SEI (Supplementary Fig. 43). This reduced Li is isolated by SEI, leading to the formation of "dead" Li and moss Li[65,66]. The "dead" Li and repeated formation of SEI give rise to low CE, accounting for why Li CE is much lower in LCE and PLHCE (Fig. 2i). The SEI with high electrical conductance and low critical field strength is more susceptible to local electric field variation, such as that induced by protuberances of Cu surface. If the local electric field strength is higher than the critical field strength of SEI, localized high electrical conduction will lead to localized Li$^+$ reduction with or on the surface of SEI, and/or localized thickening of SEI layer. The electrical behaviors of SEIs, including electrical conductance and critical field strength, account for their surface uniformity and topographical features.

Nucleation and growth of SEI are mainly based on electron tunneling model, consisting of reduction products of electrolytes formed through the reactions between the electrode and the electrolyte[2,3,27,49,67,68]. Associated with the critical thickness for electron tunneling, growth of SEI layer is self-limiting to a thickness of 2-3 nm[27,63], which is apparently far deviated from experimentally determined value of ranging from 10 nm to 50 nm[52,53,69,70]. The mechanism for further growth beyond the critical tunneling thickness of 2-3 nm remains elusive. Several mechanisms were proposed, i.e. electron diffusion through point defect like Li interstitials[71,72], solvent diffusion[73-75], electron conduction through SEI layer[73-78] and transition metal enabled electron transfer[79,80] to account for SEI growth. Further, electrons can go through SEI through point defects, such as atomic Li, especially at low voltages. The voltage dependent electron leakage mechanism has been included in many battery life models and appears to be the only one to explain some of experimental observations.[3,17,27,72-76]. The electrical properties of SEIs are determined by the microstructure and chemistry of SEI components formed by reduction and reaction of electrolyte solvent and Li salt, SEI with high content of inorganic spices shows good electrical insulation. Variation of electrolyte systems for different battery systems offers plenty room for artificial intelligence (AI), and particularly its fruitful branch known as machine learning (ML), which can be integrated to delineate the critical component and proper electrolyte chemistry to lead to ideal SEI properties, i.e. high ionic conductivity and electronic insulation, thus solving major challenges of battery research[81-84].

With our newly developed in-situ bias TEM method, for the first time, we measured the electrical properties of beam sensitive SEIs formed on the Cu and Li substrates. The results clearly reveal that behaviors of the SEI deviate from an insulator, showing voltage dependent differential electrical conductance. It is apparent that a slight variation of rate of differential electrical conductance can result in dramatic differences in SEI layer thickness and Li morphology, and consequently, the electrochemical performances of the batteries. This work resolves the long-standing puzzle as how the electrical behaviors of SEIs, as a root cause, control the electrochemical performances of rechargeable Li-based batteries. The method established here can be generally used for all types of electrochemical cells.

**Acknowledgments:** We thank Dr. Yingqing Wu in Pacific Northwest National Laboratory (PNNL) for providing $TiO_2$ sample, Dr. Chenhui Yan in University of Chicago for helpful discussions about electrical measurement. This work was supported by the Energy Efficiency and Renewable Energy, Office of Vehicle Technologies of the U.S. Department of Energy (DOE) under the Advanced Battery Materials Research (BMR) Program and the US-Germany Cooperation on Energy Storage under Contract DE-LC-000L072 (to C.W. and W.X.). P.B.B. and J.M.S. acknowledge the US-Germany Cooperation on Energy Storage under Contract DE-AC02-05CH11357, and the Assistant Secretary for Energy Efficiency and Renewable Energy, Office of Vehicle Technologies of the U.S. DOE through the BMR Program (Battery500 Consortium phase 2) under DOE Contract DE-AC05-76RL01830 from the Pacific Northwest National Laboratory (PNNL). Computational resources from the Texas A&M University High Performance Research Computing are gratefully acknowledged. The characterization work was conducted in the William R. Wiley Environmental Molecular Sciences Laboratory (EMSL), a national scientific user facility sponsored by DOE's Office of Biological and Environmental Research and located at PNNL. PNNL is operated by Battelle for the U.S. DOE under Contract DE-AC05-76RL01830. The work at the molecular foundry, Lawrence Berkeley National Laboratory (LBNL) was supported by the Office of Science, Office of Basic Energy Sciences of the U.S. DOE under Contract DE-AC02-05CH11231. G.R. and J.L. acknowledge U.S. National Institutes of Health (NIH) grants R01HL115153, R01GM104427, R01MH077303, R01DK042667.

**Author contributions:** C.W. and Y.X. conceived the project and designed the experiments with suggestion from W.X. and H.J. Y.X. collected, analyzed experimental data for in-situ TEM, cryo-TEM studies, and drafted the manuscript under the direction of C.W. and W.X. H.J. performed the electrochemical measurements. P.G. performed the AIMD calculation of bulk electrolytes. D.E.G.

**Figure Legends**

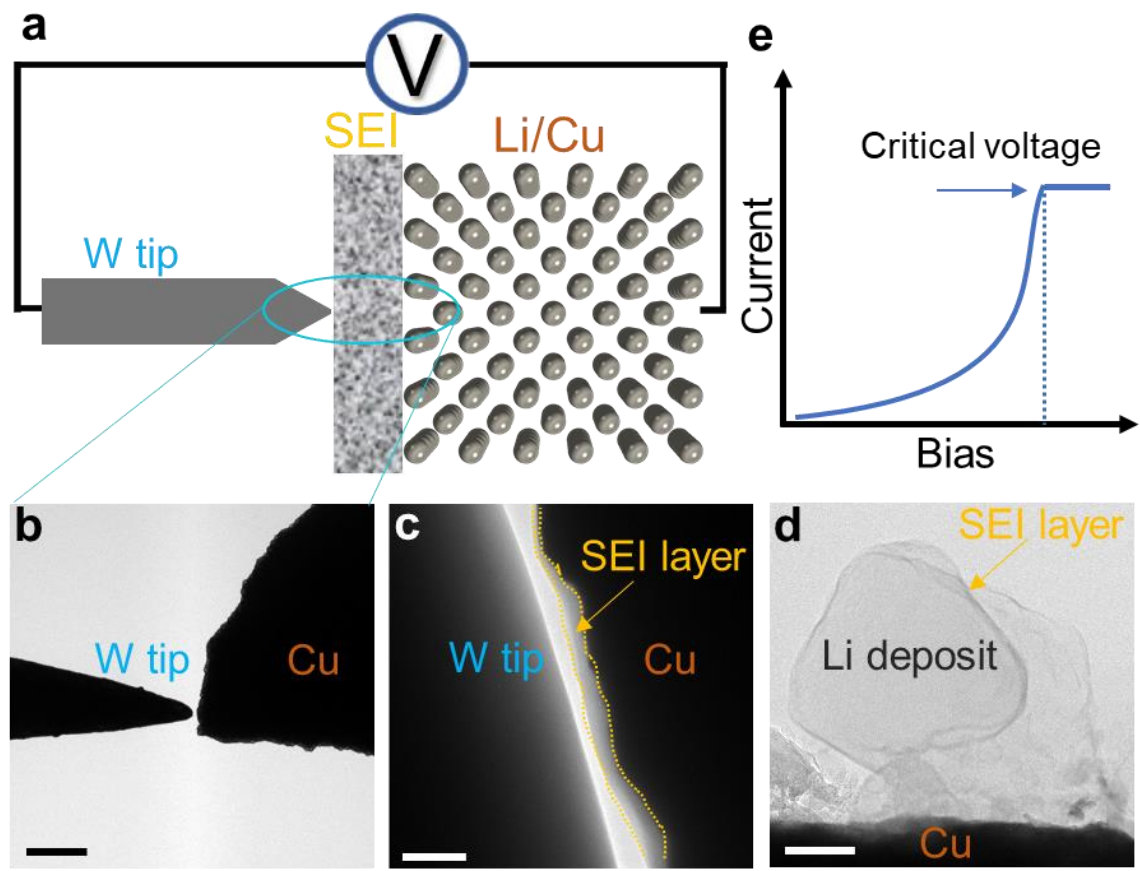

**Fig. 1: Experiment setup of preserving and measuring of SEI formed on Cu and Li by in-situ biased TEM. a**, Schematic of experiment setup. **b-d**, TEM images showing W wire with sharp tip and Cu wire with SEI layer on the surface. **e**, Typical I-V curves showing the critical voltage. Scale bars, 50 µm in **c** and 100 nm in **b** and **d**.

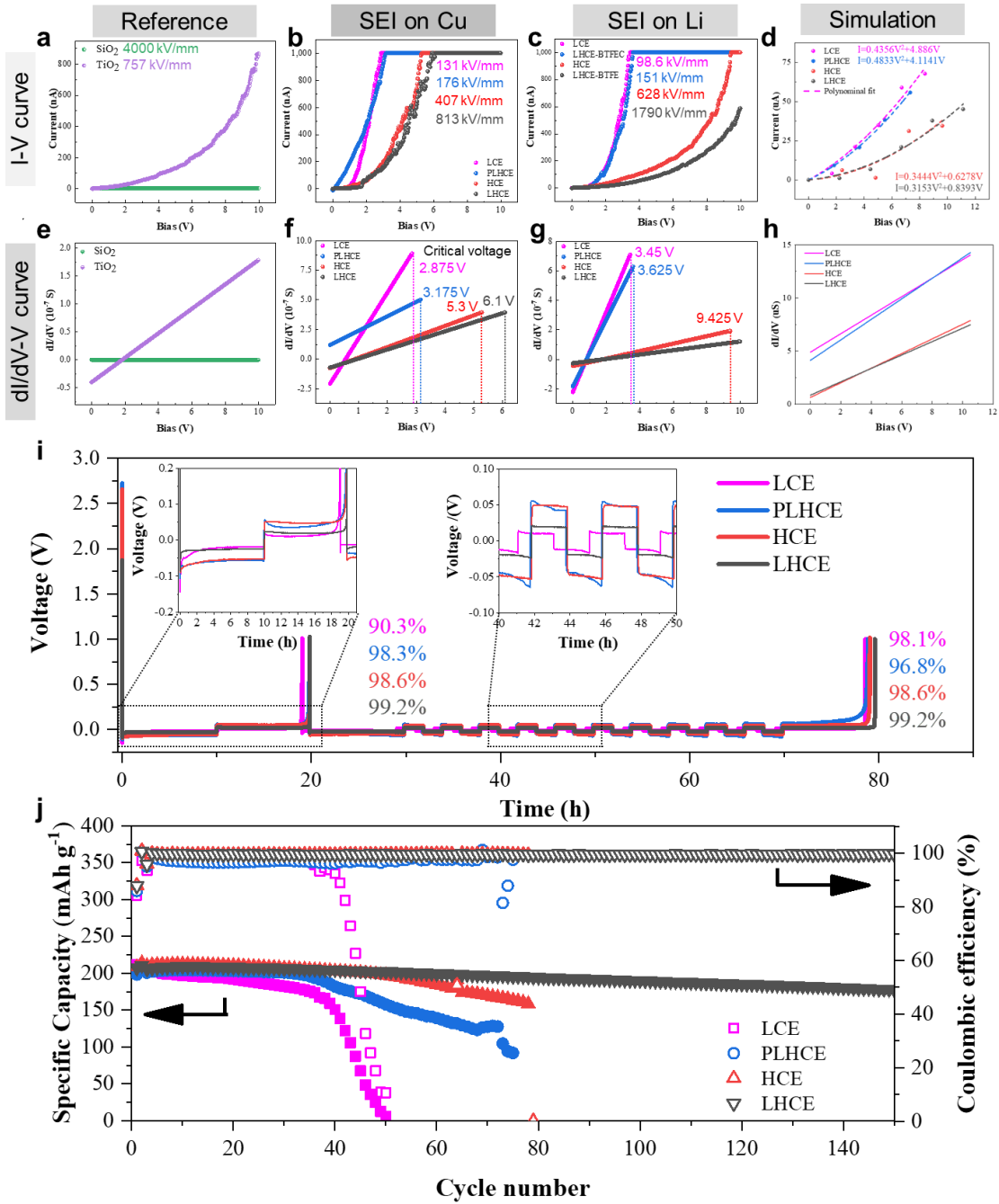

**Fig. 2: Electrical transport measurements of SEI formed on the Cu and Li in different electrolytes, and electrochemical performances of different electrolytes in Li||Cu and Li||NMC811 cells. a,** I-V curves of $SiO_2$ insulator and $TiO_2$ semiconductor. **b**, I-V curves of the SEI formed on Cu, and **c**, I-V curves of the SEI formed on Li deposits. **d,** Calculated I-V curve based on sample cell (E/A ratio is 2.79, simulation time is 253 ps). **e-h**, Differential conductance, dI/dV as function of V, derived from the I-V curves, with the critical voltage indicated. The slope of the dI/dV against V in (d-f) is termed as rate of differential conductance. **i,** CE of Li||Cu cells, and (**j**) long-term cycling stability of Li||NMC811 cells in LCE, PLHCE, HCE, and LHCE electrolytes. Inset numbers of Fig. 2**i**: initial CE and average CEs from 10 cycles.

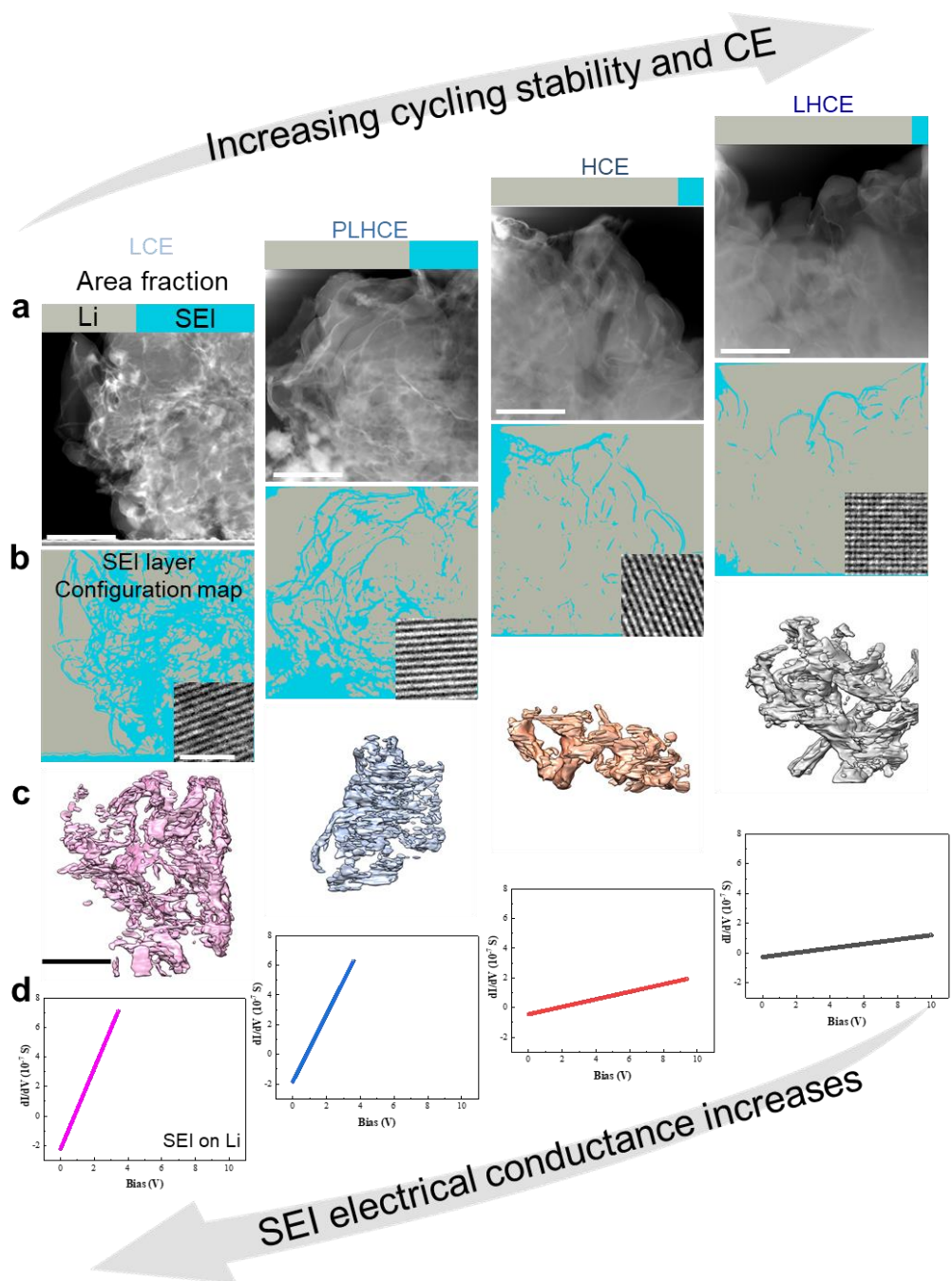

**Fig. 3: Dependence of microstructure of Li deposits on rate of differential conductance. a**, Low magnification cryo-STEM-HAADF images of Li deposits formed in LCE, PLHCE, HCE, and LHCE; **b**, SEI layer configuration maps derived from the STEM-HAADF images, the inset at each image is the high-resolution TEM images of Li deposits; **c**, 3D reconstruction of Li deposits. **d**, dI/dV-V curves of SEI on Li formed in those four electrolytes, where the slope of dI/dV as a function of V is termed as rate of differential conductance. Scale bars, 5 µm in **a** and **c**, 5 nm in the inset of **b**.

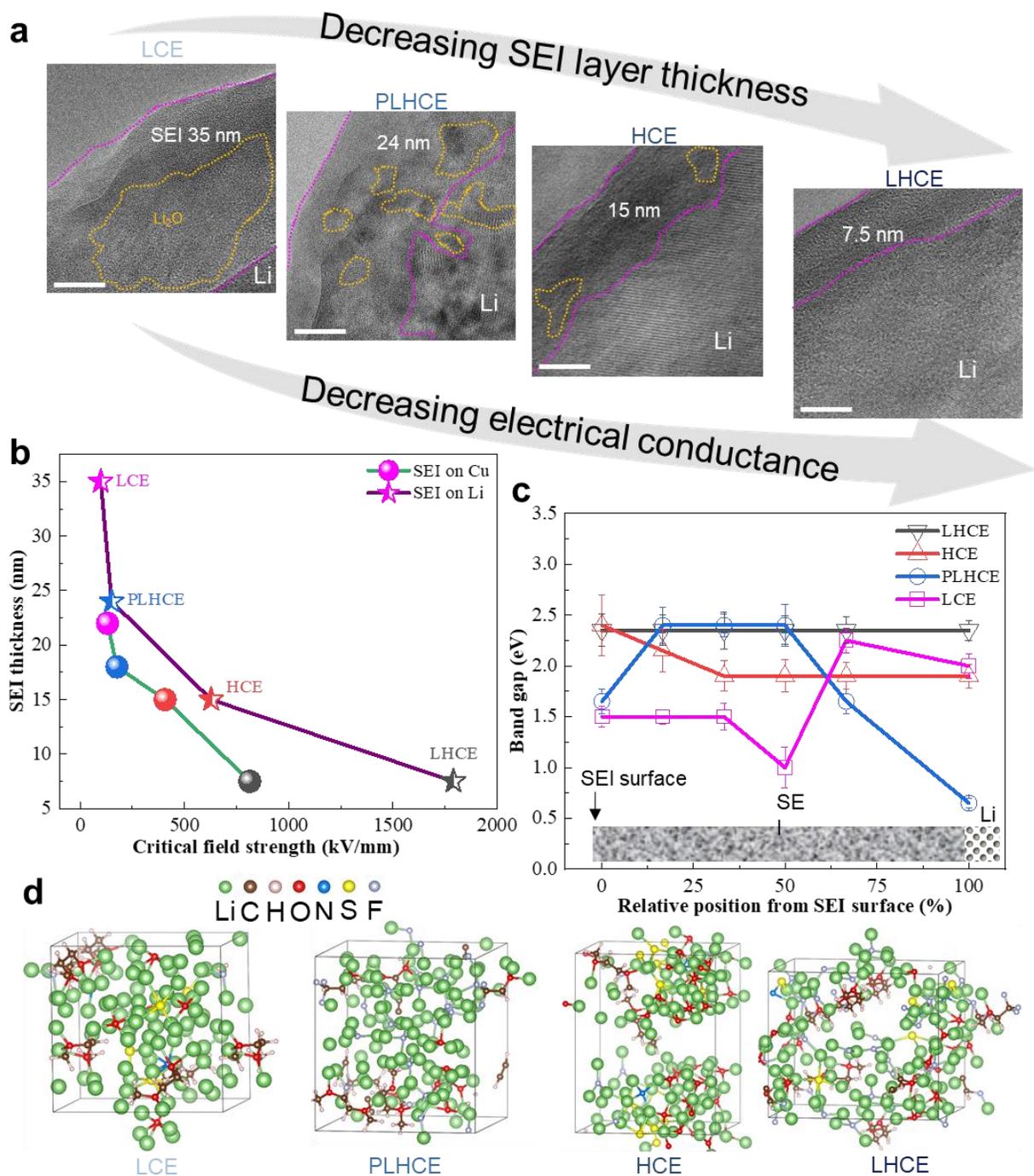

**Fig. 4: Correlation between SEI structure and its electrical property. a**, Atomic structure of SEI layers on the Li deposits formed in LCE, PLHCE, HCE, and LHCE. **b**, SEI thickness as function of the critical field strength of SEI on Cu and Li, indicating SEI layer thickness decreases with increasing critical field strength. **c**, Measured band gap of SEI layer on Li for different electrolytes, demonstrating band gap decrease from SEI surface toward Li interface. **d**, Snapshots of samples for four electrolytes reacting with Li metal in I-V curve calculations (E/A ratio is 2.79, simulation time is 253 ps). Scale bar, 5 nm in **a**.